# On the foundations of the classical relativistic theory
# of the field of an accelerated extended charge

Magomed B. Ependiev[*]

*A. A. Blagonravov Institute for Machine Science*
*of the Russian Academy of Sciences,*
*4 Maly Kharitonyevsky Pereulok, 101990 Moscow, Russia*

**Abstract.** The effect of nonzero extent of an electric charge is considered within the assumption that the structure of the charge at rest is spherically-symmetric and the current vector is linear in the acceleration. An exact expression for the electromagnetic field of the charge is obtained, which depends on the specific form of the charge distribution. We have developed the approximations which deal with the charge distribution through its low-order moments, for the case in which the particle velocity does not considerably change over the time it covers a distance of the order of its own size. We have also rigorously justified the Lorentz–Abraham–Dirac expression for the radiation friction (we have identified a more general context for this expression as well as its applicability domain). We have also studied the radiation field and demonstrated that in some cases, the radiation virtually vanishes even for large accelerations. Ways of further development of the theory have been pointed out, in order to include more general forms of the current vector (dependence of the deformation of the charge structure on the acceleration, rotation of the structure around the centre of the charge, ultrarelativistic regimes).

Many problems of the relativistic theory of microscopic particles originate from their representation as point-like particles. Renormalization procedures that are invoked to eliminate the divergences, despite their efficiency, violate the internal integrity of the theory. An account for the nonzero size of a particle is quite challenging within the quantum theory because of its statistical nature. On the other hand, the non-relativistic classical theory is able to describe an extended particle; still, to date, hardly enough attention has been paid to the relativistic generalization of such a description (perhaps, due to considerable mathematical difficulties arising in the way to it).

[*] E-mail: m010148@mail.ru.

One of the problems in which a point idealization leads to paradoxes is the problem of the interaction of a charge with its own field. This physical situation takes place, in particular, within the process of braking by radiation (radiation friction) described by the Lorentz–Abraham–Dirac formula. The traditional techniques of deriving this formula (see Ref. [1]) cannot be considered strict enough. Some approaches are based on obtaining first the non-relativistic expression and then transforming it to a relativistically-invariant form [2]. More rigorous is the Dirac's approach [3], within which the action of the field upon a charge is accounted for by taking the limit of the difference between the retarded and the advanced potentials as the charge radius tends to zero. The divergence arising within such a consideration is then eliminated using renormalization techniques (see Ref. [4]).

Within the present paper, we assume that the extent of the particle has a real physical meaning and can be described in terms of a certain internal charge distribution. This charge distribution is unknown to us yet; still, we can put forward hypotheses on its general properties and, correspondingly, construct this or that expression for the current density. Then, the approximate final results will depend on the distribution through its integral properties (total charge, low-order moments, field energy of the charge at rest, etc.).

Let us denote $x^k$ ($k = 0, 1, 2, 3$; $x^0 = ct$) the space-time coordinates of the point to evaluate the electromagnetic field in. Moreover, let $f^k(s)$ be the trajectory of the charge centre, parameterized by the space-time interval $s$ so that $ds^2 = df_k \, df^k$ (assuming a summation over repeated upper and lower indices). A four-vector $(q^0, q^1, q^2, q^3) = (q^0, \vec{q})$ will also be sometimes denoted as $q$, without specifying the vector index. The velocity vector $u = \dot{f} = df/ds$, where we have chosen the positive direction for $s$ in such a way that $u^0 > 1$ ($u^0 = (1 + \vec{u}^2)^{1/2}$). Finally, in what follows, we assume infinite integration intervals, unless the integration limits are explicitly specified.

# 1. The current density vector

An integral of the form

$$J^k(x) = \int \tilde{J}^k(x,s)ds \tag{1}$$

is a 4-vector if and only if $\tilde{J}^k(x,s)$ is also a 4-vector. For a point-like charge, Eq. (1) represents the current if $\tilde{J}^k = ecu^k\delta^{(4)}(x-f)$, where $e$ is the charge and $c$ is the speed of light. However, a replacement of the 4-dimensional delta function by its `realistic' (extended) prototype does not lead to a 4-vector anymore. In particular, a domain defined by the inequalities $|x^k - f^k| \le \delta_k$ $(k = 0, 1, 2, 3$ and $\delta_k > 0$ are certain fixed scalars), is not relativistically-invariant. In order to define a bounded domain around a space-time point in the scalar (invariant) form, one needs at least one extra vector quantity to be introduced. For an extended charge in question, the vectors are $\Delta = x - f$, $u$, $\dot{u} = \dfrac{du}{ds}$, ... ($x$ and $f$ are not "true" vectors themselves since they depend on the choice of the origin of the coordinate system). From these quantities, we can construct scalar functions

$$\tau = (u_k \Delta^k), \quad y = (\dot{u}_k \Delta^k), \quad \Delta^2 = (\Delta_k \Delta^k), \quad \sigma^2 = \tau^2 - \Delta^2 \ge 0, \dots \tag{2}$$

It is quite easy to show that the inequalities $|\tau| \le \delta_1$, $\sigma \le \delta_2$ define a bounded space-time domain for fixed positive $\delta_1$ and $\delta_2$. For a charge at rest, $\tau = ct - s$, $\sigma = r$ ($r$ is the distance between the centre of the charge and the observation point). Thus, it is natural to call $\tau$ and $\sigma$ the generalized time and distance, respectively.

If one assumes a spherically symmetric spatial structure of the charge, one may let $\tilde{J}^k = cq^k D(\tau, \sigma)$, and then, assuming that $D(\tau, \sigma)$ is a localized and even function of both arguments, search for various possible expressions for the vector $q^k$. For a simplest choice $q^k = u^k$, the continuity equation $\partial J^k / \partial x^k = 0$ is satisfied only for uniform motion. In general, however, it is necessary to account for the effect of the acceleration as well. For this purpose, let us assume a linear dependence of $q^k$ on the acceleration. Then the continuity equation yields $q^k = (1-y)u^k + \tau\dot{u}^k$ and

$$J^k = c \int [(1-y)u^k + \tau\dot{u}^k]\, D(\tau, \sigma)ds. \tag{3}$$

It is worth pointing out here that, due to the localization of the function $D$, the vector $\tilde{J}^k$ is defined up to an additive term of the form $\dfrac{d}{ds}(P^k D)$ (where $P^k$ is a polynomial in $\Delta,\ u,\ \dot u$), which does not contribute to the integral (1). For instance, it is sometimes useful to employ this arbitrariness and present Eq. (3) in the form

$$J^k = c \int [(1-y)(\tau \frac{\partial D}{\partial \tau} + 2D) - \frac{y\tau^2}{\sigma} \frac{\partial D}{\partial \sigma}]\, u^k\ ds.$$

Let us also note that

$$\int J^0(x) d^3\vec r = 4\pi c \int_0^\infty \sigma^2\, d\sigma \int d\tau D(r,\sigma) = const = ce. \qquad (4)$$

## 2. The electromagnetic field

We will search for a solution of the Maxwell equations

$$F_{,n}^{kn} = -\frac{4\pi}{c} J^k, \quad F_{kn} = A_{n,k} - A_{k,n} \quad \left((\cdots)_{,k} = \frac{\partial(\cdots)}{\partial x^k}\right), \qquad (5)$$

with $J^k$ given by Eq. (3), restricting ourselves to the retarded potentials and taking into account the Lorentz gauge condition. As a result, we obtain

$$A^k = \int u^k (A - yQ) ds',$$

$$F_{kn} = \int \left[\left(\frac{1}{\sigma}\frac{\partial A}{\partial \sigma} - y\frac{1}{\sigma}\frac{\partial Q}{\partial \sigma}\right) b_{kn} + Q a_{kn}\right] ds', \qquad (6)$$

where

$$A(\tau,\sigma) = \frac{2\pi}{\sigma} \int_0^\infty z dz \int_{|\sigma-z|}^{\sigma+z} D_1(\tau - \tau', z) d\tau',$$

$$Q = A - \frac{1}{\sigma}\frac{\partial}{\partial \sigma} \int_\tau^\infty \tau' A(\tau', \sigma)\, d\tau', \quad D_1(\tau,\sigma) = \tau \frac{\partial D}{\partial \tau} + 2D, \qquad (7)$$

$$b_{kn} = u_k \Delta_n - \Delta_k u_n, \qquad a_{kn} = u_k \dot u_n - \dot u_k u_n$$

(everywhere in the integrands, $\Delta = \Delta(s'),\ u = u(s'),\ \dots$).

Further, we will assume that the time has zero `spread' and

$$D = \delta(\tau)\mu(\sigma) \to \tilde J^k = c(1-y)u^k \delta(\tau)\mu(\sigma). \qquad (8)$$

By definition, $\mu(z)$ is an even function, thus, it can be presented in the form

$$\mu = \frac{2}{\pi} B''(z), \text{ where } (\cdots)' = \frac{\partial(\cdots)}{2z\partial z}.$$

Then, by virtue of Eq. (8), we obtain from Eq. (7) the following expressions for $A$ and $Q$,

$$A = \theta(\tau)\frac{2}{\sigma}(B'^{+} - B'^{-}), \ \ B^{+} = B(\sigma + \tau), \ \ B^{-} = B(\sigma - \tau) = B(\tau - \sigma),$$

$$Q = \theta(\tau)\left[\frac{2}{\sigma}(B'^{+} - B'^{-}) + \frac{2\tau}{\sigma^{2}}(B'^{+} + B'^{-}) - \frac{1}{\sigma^{3}}(B^{+} - B^{-})\right] \qquad (9)$$

($\theta$ is the Heaviside step function, $\theta = 0$ for $\tau < 0$, $\theta(0) = \frac{1}{2}$, and $\theta = 1$ for $\tau > 0$).

The function $\mu(\sigma)$ is a prototype of the three-dimensional delta function,

$$\mu = \frac{e}{\sigma_0^3}\Phi\left(\frac{\sigma}{\sigma_0}\right), \quad 4\pi\int\limits_{0}^{\infty}\Phi(z)z^2\,dz = 1,$$

where $\Phi(z)$ is assumed to decrease rapidly for $|z| > 1$; the parameter $\sigma_0$ can be called the `charge radius'. The localization of $\mu(\sigma)$ mentioned above implies that, in Eq. (6), the integration domain is limited by the inequality $|s' - s_0| \stackrel{<}{\sim} \sigma_0$ , where $s_0$ can be found from the equation

$$f^0(s_0) + R(s_0) = ct, \qquad R(s) = |\vec{r} - \vec{f}(s)|, \quad (x = (ct, \vec{r})). \qquad (10)$$

In fact, one can use the above equation for moderate accelerations. Let the quantity $\beta_0$ reflect the degree of stationarity of the velocity ($\frac{d}{ds} \sim \frac{1}{\beta_0}$ ). Then, in Eq. (6), we can expand all functions of $s'$ in powers of $(s' - s_0)$, provided that

$$\beta_0 \gg \sigma_0. \qquad (11)$$

In general, this technique results in quite a complicated form of the expansion for the field, because the form of the series for $\sigma(s')$ depends on the relation between $\sigma_0$ and $\sigma(s_0)$. At the same time, in the two following situations,

$$\sigma(s_0) \ll \sigma_0 \quad \text{or} \quad \sigma(s_0) \gg \sigma_0, \qquad (12)$$

the expansions are substantially simplified, and the integration is reduced to the evaluation of the `moments'

$$q_{(n)} = 4\pi\int_0^\infty \mu(\sigma)\sigma^n\,d\sigma\,, \quad n = 1,2,3,\dots \qquad (q_{(2)} = e). \qquad (13)$$

Now, in the first of the two cases in Eq. (12), we will obtain the field near the centre of the charge; in the second case, outside the charge.

***The field in the centre of the charge*** ($\vec{r} = \vec{f}(s)$ and $s_0 = s$, $\Delta(s_0) = 0$). By neglecting fourth- and higher-order derivatives of the velocity, by virtue of Eq. (9), we obtain

$$F_{kn}^{(0)} = \tfrac{2}{3} q_{(1)} a_{kn} - \tfrac{2}{3} e P_{kn}^{(0,2)} + \left( \tfrac{\lambda a_{kn}}{60} + \tfrac{1}{6} P_{kn}^{(0,3)} + \tfrac{1}{12} P_{kn}^{(1,2)} \right) q_{(3)}, \qquad (14)$$

$$\lambda = (\ddot{u}_k u^k) \geq 0, \qquad P_{kn}^{(l,m)} = u_k^{(l)} u_n^{(m)} - u_k^{(m)} u_n^{(l)}, \qquad u_k^{(m)} = \frac{d^m u_k}{ds^m}.$$

The quantity $q_{(1)}$ is divergent for $\sigma_0 \to 0$ if $\mu$ is a sign-definite function. On the other hand, if the latter changes sign, $q_{(1)}$ may vanish (for instance, $q_{(1)} \equiv 0$ for $\mu(\sigma) \sim \left( \frac{\sigma^2}{\sigma_0^2} - 1 \right) \exp(-\frac{\sigma^2}{\sigma_0^2})$ ).

***The field outside the charge.*** In the $\sigma(s_0) \gg \sigma_0$ case chosen in Eq. (9), we can neglect the contributions of the functions $B^+$ and $B'^+$. Now, the quantity $s_0$ is a function of $ct$ and $\vec{r}$ which is determined by Eq. (10). The integrals in Eq. (6) can be expanded into power series in the small parameters $\sigma_0/\beta_0$ and $\sigma_0/\sigma(s_0)$, up to arbitrary accuracy. By neglecting the terms of the order of $(\sigma_0/\sigma(s_0))^4$ and third- and higher-order derivatives of the velocity, we obtain

$$F_{kn}^{(1)} = e \left[ \frac{(y_0 - 1)}{a_0^3} b_{kn}^{(0)} - \frac{1}{a_0^2} b_{kn}^{(1)} \right] + \left[ \frac{y_0(y_0 + 1)}{a_0^5} b_{kn}^{(0)} + \frac{y_0}{2 a_0^4} b_{kn}^{(1)} + \frac{a_{kn}(s_0)}{6 a_0^3} - \frac{b_{kn}^{(2)}}{6 a_0^3} \right] q_{(4)}, \qquad (15)$$

where

$$a_0 = \sigma(s_0) = \left( u_k(s_0) \Delta^k(s_0) \right), \qquad y_0 = \left( \dot{u}_k(s_0) \Delta^k(s_0) \right),$$

$$b_{kn}^{(m)} = [u_k^{(m)} \Delta_n - \Delta_k u_n^{(m)}]_{s=s_0}.$$

The first term in Eq. (15) gives the field of a point change described by the well-known Liénard–Wiechert potentials. This does not imply, however, that the second term introduces only (small) corrections to the first one. Depending on the relation between the values of $\beta_0$ and $a_0$, the orders of magnitude of the two terms may vary.

### 3. Interaction of the charge with its own field

If the field inside of the charge were homogeneous enough, the force acting on the charge from the field could be defined as the Lorentz force $g_k = \frac{e}{c^2} F_{kn}^{(0)} u^n$. However, this conjecture is adequate only within those approaches in which the size of the charge is set to zero and renormalization is employed in order to eliminate the divergences. In contrast, we will work within the assumption that the charge has a `true' extended structure and, having already figured out the form of the field created by such a charge, will follow the method of the classical theory.

The part of the Lagrangian containing the field of the charge, is totally determined by the space-time trajectory of the charge. The equations of motion for the charge are derived by equating the variation of the action to zero, with respect to variations of the trajectory. Note that $u_k \delta u^k = 0$, thus, the functional derivative with respect to the trajectory should include a term $\frac{d}{ds}(Pu^k)$, with $P$ chosen so that $g_k u^k = 0$ ($g_k$ is the force to be determined). Assuming that the current is expressible in the form (8), we arrive at the general expression for the force

$$g_k(s) = \frac{1}{c^3} \int F_{kn}(x+f)\tilde{J}^n(x+f,s)d^4x + \dot{C}_k, \qquad (16)$$

$$C_k = \frac{1}{c^2} \int [yA_k - yA_n u^n u_k - A_n \dot{u}^n x_k] \, \delta(\tau)\mu(\sigma)d^4x,$$

Where $\tau = (u_k x^k)$, $y = (\dot{u}_k x^k)$, $\sigma^2 = \tau^2 - x_k x^k$, $A_k = A_k(x+f)$ (we have shifted the integration variables, $x \to x - f$). Further, we will also obtain the expression for the force acting on the charge from the external field, which will be quite analogous to Eq. (16). Here, the only correction should be taken into account, the external field should not depend on the trajectory of the charge. If the external field varies slowly at the length scale of the charge size, then the external force is the Lorentz force. Otherwise, this force should be calculated from an expression analogous to Eq. (16).

Quite naturally, in the general case, a sufficient analysis of Eq. (16) requires the information on the specific form of the function $\mu(\sigma)$. Still, once the condition (11) is met, we are able to obtain a rapidly convergent series, with its terms depending on $\mu(\sigma)$ through its integral properties. Moreover, in this case, the second term in Eq. (16) is a contribution proportional to at least forth-order derivatives of the velocity. By neglecting such contributions, we obtain from Eq. (16)

$$g_k = -M\dot{u}_k + \frac{2e^2}{3c^2}(\ddot{u}_k - \lambda u_k) + \alpha_0 \frac{1}{3c^2}\frac{d}{ds}(\frac{3}{2}\lambda u_k - \ddot{u}_k), \qquad (17)$$

$$M = \frac{8}{c^2}\int_0^\infty (B'(\sigma))^2 d\sigma, \quad \alpha_0 = -16\int BB' d\sigma = \iint |\vec{r} - \vec{q}|\mu(r)\mu(q)d\vec{r}d\vec{q}. \ (18)$$

For a charge at rest, the field strengths and the total energy of the field read

$$\vec{E} = -\frac{4\vec{r}}{r^2}\left[\frac{1}{r}\int_0^r B'(\sigma)d\sigma - B'(r)\right], \quad \vec{H} = 0, \qquad (19)$$

$$W = \frac{1}{8\pi}\int \vec{E}^2 d^3\vec{r} = 8\int_0^\infty (B')^2 d\sigma = Mc^2. \qquad (20)$$

Thus, we can realistically refer to the quantity $M$ in Eq. (17) as to the `rest' mass of the field and, by moving the first term in Eq. (17) to the left side of the equation of motion for the charge, operate with the total mass of the charge and the field (in fact, today, the latter is the only quantity we possess information on!). The second term in Eq. (17) coincides with the well-known expression for the radiation friction. The correction to the first term in Eq. (17), namely, the third term, is subject to an additional quadratic suppression in $\sigma_0/\beta_0$, compared to the first term.

## 4. The radiation

For the terms of the order of $1/r$ in the $r \to \infty$ limit, we obtain

$$\vec{E} = -\frac{2}{r}\int \dot{\vec{b}}\frac{B'(z)}{\varepsilon}ds, \quad \vec{H} = [\vec{n}\ \vec{E}], \quad \vec{n} = \frac{\vec{r}}{r}, \quad \vec{E}\vec{n} = 0, \qquad (21)$$

$$\vec{b} = \frac{1}{\varepsilon}[u_0\vec{n} - \vec{u}], \quad \varepsilon = u_0 - \vec{u}\vec{n}, \quad z = \frac{1}{\varepsilon}\left[r - ct + f_0(s) - (\vec{n}\vec{f}(s))\right].$$

If the condition (11) holds, we can make an expansion in Eq. (21) in powers of $(s - s_0)$, where $s_0$ is determined by equation (10) now taking the form $f_0(s_0) - (\vec{n}\vec{f}(s_0)) = ct - r$. By neglecting fourth- and higher-order derivatives of the velocity, we obtain

$$\vec{E} = \frac{e}{rc}\frac{\partial \vec{b}}{\partial t} + \frac{q_{(4)}}{6rc^3}\frac{\partial^2}{\partial t^2}(\varepsilon^2 \frac{\partial}{\partial t}\vec{b}), \qquad (22)$$

where $u_k = u_k(s_0)$ and $s_0$ is itself a function of $(ct - r)$ and $\vec{n}$.

Of much current interest are the cases in which the condition (11) is not met. Here, depending on the form of the distribution $\mu(\sigma)$ and the motion dynamics of the charge, various regimes might be possible, up to the case of virtually vanishing radiation. For example, let us consider a periodic motion. By changing the integration variable from $s$ to $t' = f^0(s)/c$ in Eq. (21), we obtain

$$\vec{E} = \frac{1}{r\gamma_0} \int_0^T \left( \frac{\partial \vec{b}}{\partial t'} \right) \left[ \sum_{m=1}^{\infty} \bar{\mu} \left( \frac{2\pi m \varepsilon}{\gamma_0} \right) \cos \left( \frac{2\pi m}{\gamma_0} z \right) \right] dt', \qquad (23)$$

where $T$ is the oscillation period, $\gamma_0 = cT$, $z = \left( r - ct + ct' - \left( \vec{n}\vec{f} \right) \right)/\varepsilon$, and

$$\bar{\mu}(p) = \frac{1}{2} \int e^{-i\vec{p}\vec{r}} \, \mu(r) d^3\vec{r} \, .$$

If $\mu \sim \exp(-\sigma^2/\sigma_0^2)$, then, at $\gamma_0 \cong \frac{\sigma_0}{2}$, we have $\left| \vec{E} \right| \sim \frac{e}{r\gamma_0} \times 10^{-16}$ and the radiation virtually vanishes. Let us also point out that Eqs. (21)–(23) (in fact, as well as all the above results) are also applicable for neutral objects, for which $q_{(2)} = e = 0$ (of course, if the structure of the `charge' contains charged elements).

## 5. The time spread

In contrast to the classical extent of the charge, the notion of the `time spread' is less transparent and requires identification of its physical meaning within the classical theory. We set this problem aside and, in the present paper, consider what will change if one takes into account the time spread.

Let us assume the following generalization of (8) to take place in (3),

$$D = V(\tau) \, \mu(\sigma), \quad \int V(\tau) d\tau = 1, \quad V(\tau) = V(-\tau), \quad V(\tau)|_{|\tau| \gg ct_0} \to 0, \quad (24)$$

where $t_0$ describes the time spread. In the $ct_0 \ll \beta_0$ regime, we arrive at small corrections to the results following from (11). But, in the $ct_0 \gtrsim \beta_0$ regime, the situation does change. The acceleration might be small compared with the `charge radius' (i.e., the condition (11) might hold), nevertheless, it is no more legal to expand the field into a series in the derivatives of the velocity. At the same time, the case in question does strongly differ physically from the cases in which (11) does not hold.

The spatial structure of the charge is determined by internal forces that are unknown to us. And, in principle, it may happen, for instance, that the real size of the electron is comparable with its classical radius. Then the realization of the $\sigma_0 \gtrsim \beta_0$ case would require practically unattainable accelerations. On the other hand, the time spread may be essentially controlled by the external conditions. In particular, for stationary processes (uniform motion, steady-state oscillations or rotation, etc.), we may expect large values of the quantity $t_0$.

All the above tells us that it is worth considering the field in those cases in which the `time spread' is much greater than the spatial one,

$$ct_0 \gg \sigma_0 \qquad (25)$$

In these cases, it is quite straightforward to derive from Eqs. (6), (7) the expansion for the field and the analogues of the other above results. From Eq. (7), one obtains for $r \rightarrow \infty$

$$A = \frac{e}{\sigma} \, G(\sigma - \tau), \quad Q = \frac{e}{\sigma^2} \, (\sigma - \tau)G(\sigma - \tau), \quad G(\tau) = \tau\frac{dV}{d\tau} + 2V. \quad (26)$$

This leads to the radiation field

$$\vec{E} = \frac{e}{r} \int \dot{\vec{b}} \, \frac{G(z)}{\varepsilon} \, ds, \qquad \vec{H} = [\vec{n} \; \vec{E}], \qquad (27)$$

which differs from Eq. (21) only by a replacement of $2B'(z)$ by $-eG(z)$. Moreover, now, in the approximate expression (22),

$$q_{(4)} = 3e \int \tau^2 \, G(\tau)d\tau \qquad (28)$$

and the representation (23) contains the spectrum of the function $G(\tau)$,

$$\bar{\mu}(p) \quad \rightarrow \quad \bar{G}(p) = e \int cos(p\tau) \, G(\tau)d\tau. \qquad (29)$$

In the case under consideration, the `moment' $q_{(4)}$ in Eq. (22) appears to be much greater than before and the contribution of the second term takes effect at smaller accelerations. Quite similarly, in the case of a periodic motion, the radiation field (23) almost vanishes at much smaller oscillation frequencies.

## 6. Structure deformation by acceleration

The representation (3) of the current vector was based on the assumption that the deformation of the charge structure is caused only by the velocity (the Lorentz contraction). In the case (11), such an assumption is indeed justified. However, we cannot discard the possibility for the acceleration to affect the deformation as well. We can account for this effect by introducing the dependence of the charge distribution on $y = (\dot{u}_k\Delta^k)$. Then, the choice of possible forms of the current vector becomes significantly richer (especially because one has to introduce a parameter describing the `elasticity' of the structure). We will dwell here on the two simplest cases.

In the first case, let $\tilde{J}_k = q_k D(\tau - gy, \sigma)$, with $g$ being a scalar function of $s$ (possibly a constant). This leads to the `deformation' of time. Assuming a linear dependence of $q_k$ on $\ddot{u}$, we obtain

$$J_k = \int [(1 - y + \dot{g}\, y + g\dot{y} - \lambda\tau g)u_k + \tau\dot{u}_k]D(\tau - gy, \sigma)ds. \qquad (30)$$

We have already mentioned above that the time spread requires an analysis of its physical meaning. The same is required for the time `shift' featuring in Eq. (30), and thus we leave the case which leads to Eq. (30) beyond our further consideration.

More transparent is the acceleration-induced deformation of the spatial structure of the charge. Let us account for this deformation by adopting the representation

$$\tilde{J}_k = q_k D(\tau, \sigma_1), \qquad \sigma_1 = \sqrt{\sigma^2 + w_0 y^2}, \qquad w_0 = const. \qquad (31)$$

The vector $q_k$ is determined by the continuity equation, yielding

$$J_k = \int [(1 - y)\gamma u_k + \frac{(\tau + w_0\dot{y})}{\gamma}\dot{u}_k]D(\tau, \sigma_1)ds, \qquad (32)$$

$$\gamma = \sqrt{1 + w_0\lambda}, \qquad \lambda = (\ddot{u}_k\, u^k).$$

A specific role in Eq. (32) is played by the constant $w_0$ (having the dimension of length squared) which describes the `elasticity' property of the structure (for an `absolutely rigid' structure, $w_0 = 0$ and (32) reduces to (3)). Depending on the sign of this `elasticity constant', the structures can be classified as stable and unstable.

If $w_0 < 0$, then, for such accelerations that $\lambda w_0 \cong -1$, Eq. (32) loses its physical meaning (for instance, in the case of motion along a straight line, the structure transforms into an infinite `string'). This means that such structure is destroyed by high accelerations. If $w_0 > 0$, no such singularities arise and the structure shrinks along the direction of the acceleration (of course, only to a degree `permitted' by the internal forces). Obviously, the representation (32) also has its applicability domain. Indeed, while $|w_0| \leq \sigma_0^2$, we can consider large accelerations, at which the condition (11) is violated. But if $|w_0| \gg \sigma_0^2$, it is quite possible that the applicability domain is itself determined by the condition (11).

The technique for the analysis of the field created by the current (32) is quite analogous to that discussed above. If $|w_0| \leq \sigma_0^2$, then, provided the condition (11) is met, the `elasticity' property enters the terms that are proportional to the third- and higher-order derivatives of the velocity. We will not quote these results here (they are quite lengthy) and confine ourselves to the expression for the radiation field.

In the limit $r \to \infty$, within the representation $D = \delta(\tau)\mu(\sigma_1)$, we obtain

$$F_{ik} = -\frac{2}{r} \int \left\{ \frac{B'}{N} \frac{d}{ds} \left( \frac{[n,u]_{ik}}{\varepsilon} \right) + \left( \frac{d}{dz}(B'z) \right) \frac{w_0 y_1}{\varepsilon \gamma^2 N^3} [n,m]_{ik} \right\} ds, \quad (33)$$

where

$$B' = -\pi \int_z^\infty \sigma\mu(\sigma)d\sigma, \qquad z = [ct - r - f^0(s) + (\vec{f}\,\vec{n})]/N,$$

$$N = \frac{1}{\gamma}\sqrt{(\gamma^2\varepsilon^2 - w_0\dot{\varepsilon}^2)}, \qquad y_1 = \lambda\varepsilon + \frac{w_0}{2\gamma^2}\dot{\lambda}\dot{\varepsilon} - \ddot{\varepsilon}, \quad m_k = \varepsilon\dot{u}_k - \dot{\varepsilon}u_k,$$

$$[a,b]_{ik} = a_i b_k - a_k b_{i,} \quad n^k = (1,\vec{n}), \quad \vec{n} = \vec{r}/r.$$

It is quite obvious that Eq. (33) provides more freedom for searching the situations in which the radiation vanishes, than the freedom offered by Eq. (21) (Eq. (33) reduces to Eq. (21) at $w_0 = 0$).

Within the analysis presented above, we have considered but a few options for the current vector. In particular, we have left beyond the cases involving not only the motion of the charge as a whole but also some internal dynamics, such as the rotation of the structure around the centre of the charge (i.e., the spin). Moreover, ultrarelativistic cases require a specific treatment (since our results converge slower for velocities close to the speed of light). This all indicates the existence of a wide edge for further development of the theory. This way could, perhaps, open up the possibility to `reconcile' the classical physics with some quantum phenomena. One might also expect that certain role in this reconciliation might be played by the `time spread', which is puzzling within the classical theory.

In the present paper, we did not raise a question of the nature of the external forces that make the particle move with acceleration, assuming that large acceleration may result not only from the action of classical electromagnetic forces, but also due to fluctuations that are typical for microscopic processes (some fluctuations may lead to short but large accelerations).

   In conclusion, let us list the meanings of some of the physical constants we used in the order-of-magnitude inequalities,

   $\sigma_0$ (dimension of length) – the assumed spatial extent of the particle;

   $t_0$ (dimension of time) –  the degree of the time spread;

   $\beta_0$ (dimension of length) – the space-time interval over which a considerable change of the velocity occurs (the degree of the stationarity of the velocity);

   $w_0$ (dimension of length squared) – a parameter which describes the `elasticity' properties of the particle's internal structure.